# The Role of Excitatory Parvalbumin-positive Neurons in the Tectofugal Pathway of Pigeon *(Columba livia)* Hierarchical Visual Processing

Shan Lu[1]¶, Xiaoteng Zhang[2,3]¶, Yueyang Cang[2,3], Shihao Pan[2,3], Yanyan Peng[4], Xinwei Li[1], Shaoju Zeng[1]*, Yingjie Zhu[5]*, Li Shi[2,3]*

[1]Beijing Key Laboratory of Gene Resource and Molecular Development, College of Life Sciences, Beijing Normal University, Beijing, China

[2]Department of Automation, Tsinghua University, Beijing, China

[3]Institute for Brain and Cognitive Sciences, Tsinghua University, Beijing, China

[4]Henan Key Laboratory of Brain Science and Brain-Computer Interface Technology, School of Electrical and Information Engineering, Zhengzhou University, Zhengzhou, China

[5]Shenzhen Key Laboratory of Drug Addiction, The Brain Cognition and Brain Disease Institute, Shenzhen Institutes of Advanced Technology, Chinese Academy of Sciences, Shenzhen, China

¶ These authors contributed equally to this work.

* Corresponding author. E-mail: 95034@maill.bnu.edu.cn, zhuyingjie@siat.ac.cn, shili@tsinghua.edu.cn

## Abstract

The visual systems of birds and mammals exhibit remarkable organizational similarities: the dorsal ventricular ridge (DVR) demonstrates a columnar microcircuitry that parallels the cortical architecture observed in mammals. However, the specific neuronal subtypes involved and their functional roles in pigeon hierarchical visual processing remain unclear. This study investigates the role of excitatory parvalbumin (PV+) neurons within the Ento-MVL (entoallium- mesopallium venterolaterale) circuit of pigeons underlying hierarchical moving target recognition. Electrophysiological recordings and immunofluorescence staining reveal that excitatory PV+ neurons originating from the entopallial internal (Ei) predominantly modulate MVL responses to varying visual stimuli. Using a heterochronous-speed recurrent neural network (HS-RNN) model, we further validated these dynamics, replicating the rapid adaptation of the Ento-MVL circuit to moving visual targets. The findings suggest that the fast-spiking and excitatory properties of PV+ neurons enable rapid processing of motion-related information within the Ento-MVL circuit. Our results elucidate the functional role of excitatory PV+ neurons in hierarchical information processing under the columnar organization of the visual DVR and underscore the convergent neural processing strategies shared by avian and mammalian visual systems.

## Introduction

Birds have evolved a highly sensitive visual system for safe and efficient flight, with specialized motion detection and object recognition capabilities(Chaib et al., 2021).This system is crucial for obstacle avoidance, target capture, and navigation during flight. There are certain similarities in the visual information processing mechanisms between birds and mammals(Briscoe et al., 2018).This is evident in the column-like local circuit patterns found in the dorsal ventricular ridge (DVR) of birds, which is reminiscent of the neuronal fiber orientation in the mammalian cortex(Stacho et al., 2020). This similarity could be a result of convergent evolution between birds and mammals. The visual nuclei within the DVR are part of the tectofugal pathway, involved in the parallel processing of shape, color, and 3-D object motion/looming information, similar to the processing mechanisms found in the mammalian ventral stream, associated with the emergence of anatomically distinct parallel processing streams for form, color, and 3-D object moving/looming information(Aflalo and Graziano, 2011; Chatterjee and Callaway, 2003; Clark and Colombo, 2020; Gur and Snodderly, 2007).Within the visual region of the DVR, Ento (entopallium) and MVL (mesopallium ventrolaterale) are situated along the same vertical axis, with the Ento located ventrally and the MVL dorsally, forming an orthogonal laminar organization(Ahumada-Galleguillos et al., 2015; Fernández et al., 2020; Stacho et al., 2020).Functionally, Ento and MVL exhibit a hierarchical information processing architecture, analogous to that found in mammals. Ento is responsible for processing basic visual features such as color and shape, whereas the MVL is engaged in higher-order visual tasks, including object recognition and categorization(Anderson et al., 2020; Azizi et al., 2019; Clark et al., 2022; D'souza et al., 2022; Lu et al., 2018; Niu et al., 2024; Zhu et al., 2025).In terms of neural connectivity, Ento is analogous to layer 4 of the mammalian neocortex, which primarily receives thalamic input, while the MVL corresponds to layers II/III, known for intracortical integration and processing(Atoji and Karim, 2014; Briscoe et al., 2018; Fernández et al., 2020).Previous experimental studies have demonstrated bidirectional projections between the Ento and MVL(Ahumada-Galleguillos et al., 2015; Atoji and Wild, 2012; Fernández et al., 2020); however, the specific neuronal subtypes involved and their functional connectivity patterns that support this hierarchical organization remain to be elucidated.

Parvalbumin-expressing (PV+) neurons play critical roles in visual processing, sensorimotor integration, and higher-order cognitive behaviors in mammals(Runyan et al., 2010; Sohal et al., 2009). In birds, PV+ neurons in the isthmi pars magnocellularis (IMC) and HVC (high vocal center) are involved in spatial attention control and song learning (a form of motor learning), respectively, functions that parallel those of PV+ neurons in mammals.(Mahajan and Mysore, 2022; Zuschratter et al., 1985) Interestingly, in addition to the classical inhibitory PV+ neurons, mammals also possess excitatory PV+ neurons, which are distributed across multiple brain regions, including the deep layers of the visual cortex, the superior colliculus (SC), the red nucleus (RN), and the entopeduncular nucleus (EP)(Rizzi et al., 2019; Shang et al., 2015; Wallace et al., 2017).Functionally, excitatory PV neurons contribute to triggering stereotyped fear responses via the visual pathway, participate in the learning of fine motor skills such as grasping, and play essential roles in the modulation of emotion, motivation, and reward-related processes, forming part of the vertebrate midbrain network for the control of spatial attention(Rizzi et al., 2019; Shang et al., 2018; Shang et al., 2015; Wallace et al., 2017). In the entopallial internal (Ei), subregion of Ento, a dense population of PV neurons has been observed(Krützfeldt and Wild, 2005). However, the specific functional roles of PV neurons within the Ento-MVL visual DVR region in birds remain largely unknown.

This study found that the density of excitatory neurons in the visual regions of DVR in birds is significantly higher than that of inhibitory neurons, displaying structural features similar to the distribution of excitatory and inhibitory neurons in the VISp (primary visual cortex) of mice. Additionally, the DVR region exhibits a hierarchical processing of visual information from low to high level, especially after the MVL region receives input from Ei, becoming involved in more complex visual processing. In this process, excitatory PV+ neurons play a crucial role. The E-MVL circuit demonstrates strong adaptability in processing motion-related visual information. This conclusion was further validated by asynchronous recurrent neural network (RNN) simulations, which showed that excitatory PV+ neurons enable the pigeon E-MVL circuit to rapidly respond to moving targets.

Results

This study initially used herpes simplex virus (HSV) to investigate the neural circuitry between Ento and MVL regions in the pigeon visual DVR area. HSV-H129 transmits to the 2nd, 3rd, and 4th order neurons at about 45, 55, and 65h after virus injection(Tang et al., 2016; Yang et al., 2021), our samples were collected at 46, 56 and 72hours post-injection, and were compared with CTB (Choleratoxin Subunit B) retrograde tracing and cytochrome oxidase (CO) labeled regions. Immunofluorescence staining was used to distinguish neuron types and explore the similarities and differences in microcircuit organization between pigeon and mice (Figure 1a, 1b).

To study the neural response properties of Ento and MVL, multi-channel electrode arrays (MEA) were used to perform electrophysiological recordings from the left MVL and left Ei regions of 18 pigeons (Figure 1d), considering the left hemisphere dominance of Ento and the right eye preference(Verhaal et al., 2012). In the experiment, the pigeons' heads were fixed, the left eye was occluded, and the right eye viewed visual stimuli (Figure 1e). After the experiment, the accuracy of the target site implantation was confirmed using CO staining (Supplementary Figure 11). To comprehensively explore the mechanism of information processing between Ento and MVL, a variety of stimuli were used, including 6 colors, 8 motion directions, 10 spatial frequencies of gratings stimuli, and 8 categories of natural image stimuli (Figure 1c, supplementary materials). The electrophysiological signals were processed using Kilosort for spike sorting(Pachitariu et al., 2016), isolating high-quality neurons from the Ei and MVL regions (Figure 1f).

Electrophysiological analysis and single-cell RNA sequencing results indicate that bird NS (narrow spiking) neurons align with the waveform pattern of mammalian PV neurons(Kvitsiani et al., 2013; Schneider and Woolley, 2013; Yanagihara and Yazaki-Sugiyama, 2016). Based on this, it is speculated that NS waveform neurons in the pigeon brain may correspond to the PV neuron population. Using principal component analysis (PCA) and k-means clustering, the neurons were classified into three types: Broad Spiking (B), Intermediate Spiking (IM), and Narrow Spiking (N), with their full width at half height (FWHH) being $0.22\pm0.003$ ms, $0.25\pm0.003$ ms, and $0.34\pm0.007$ ms, respectively. Significant differences ($p<0.001$) were found in afterhyperpolarization (AHP), and resting firing rate (Figure 1g-i, Supplementary Figure 12b).

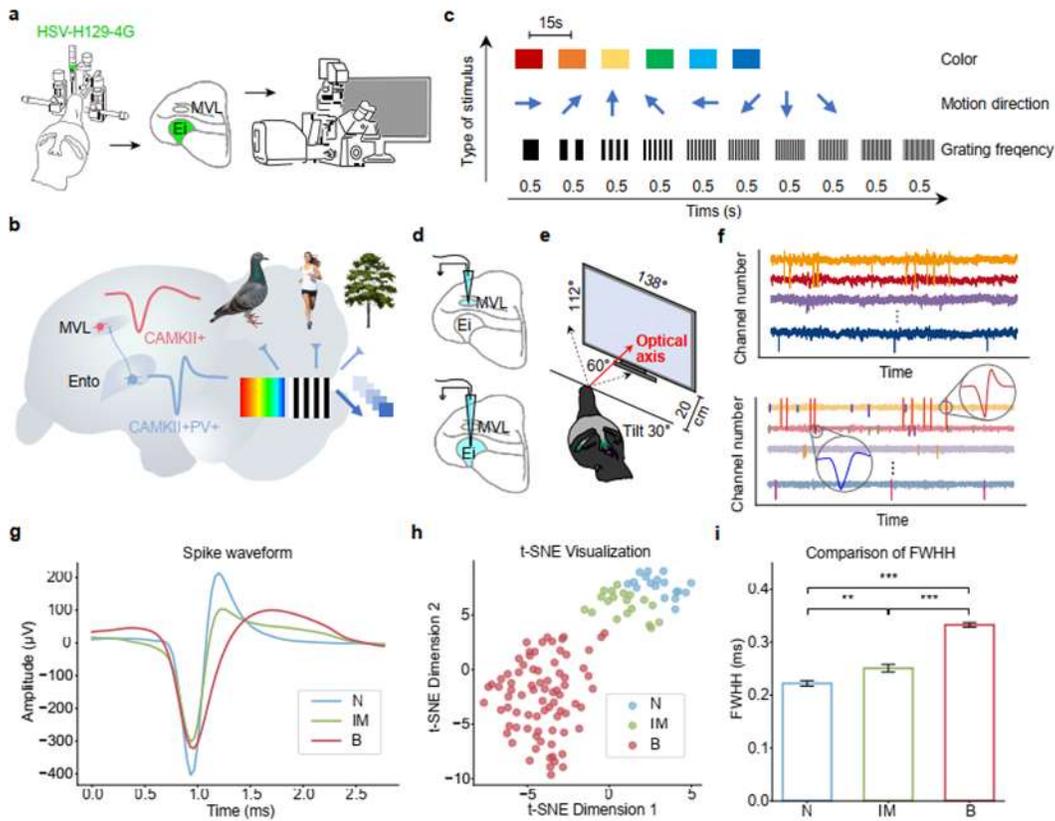

**Fig. 1. Experimental Setup. a**, Schematic of the procedure used to examine the connectivity between Ento and MVL using HSV-H129-4G or other tracers. The process includes targeted tracer injection, immunofluorescence or cytochrome oxidase staining. **b,** Schematic diagram of hierarchical information processing of Pigeon ento-mvl neural circuits and schematic diagram of natural image classification paradigms**. c,** Electrophysiological recording paradigm. Stimuli consisted of six colors, eight motion directions, and ten spatial frequencies of drifting gratings. Each stimulus was presented for 500 ms with a 15-second inter-stimulus interval. **d**, Schematic diagram of electrophysiological recording sites. **e**, Diagram of the experimental setup showing the stimulus screen positioned 20 cm from the pigeon's right eye and angled 30° relative to the body axis. **f**, Electrophysiological data acquisition and preprocessing workflow, including spike sorting from raw recordings. **g**, Average spike waveforms of three neuron types recorded in the MVL and Ei regions. **h**, Visualization of spike clustering results. **i**, Full width at half height (FWHM) of the spike waveforms for the three neuronal types (NS: $0.22 \pm 0.003$ ms; IMS: $0.25 \pm 0.003$ ms; BS: $0.34 \pm 0.007$ ms; Wilcoxon rank-sum test, **$p < 0.01$, ***$p < 0.001$, cc****$p < 0.0001$).

**1. Hierarchical Organization of Neuronal Distribution and Visual Information Processing in the DVR Region**

In the mammalian cortex, the number of excitatory neurons is significantly higher than that of inhibitory neurons(Alreja et al., 2022). To investigate the cytoarchitectural organization of the DVR subregions (MVL, MVLv, NI, Ex, and Ei) and their laminar correspondences to the mammalian neocortex, we categorized and analyzed the distribution of neuron types (Supplementary Figures 1b, d) using excitatory neuron marker CaMKII, inhibitory neuron marker GABA, and neuron marker parvalbumin. Our results indicated that, similar to mice, the density of excitatory neurons was higher than that of inhibitory neurons in the pigeon visual DVR region (the cell density of CaMKII in mice was 1020 cells/mm², GABA 211.4 cells/mm², t-test p<0.0001, in pigeons, CaMKII density was 347.4 cells/mm², GABA density was 110.6 cells/mm², t-test p<0.0001) (Figure 2b). In the mice VISp region, the density of CaMKII+ neurons was highest in layer VI (one-way ANOVA, p<0.0001, P=0.0078), and the density of PV+ neurons was highest in layer IV (one-way ANOVA, p<0.0001). In the pigeon visual DVR region, the MVL region had relatively high CaMKII+ neurons density (409.2 cells/mm²), and PV+ neurons density is highest in Ei (326.8 cells/mm²) (Figure 2b, Supplementary Figures 1b, d).

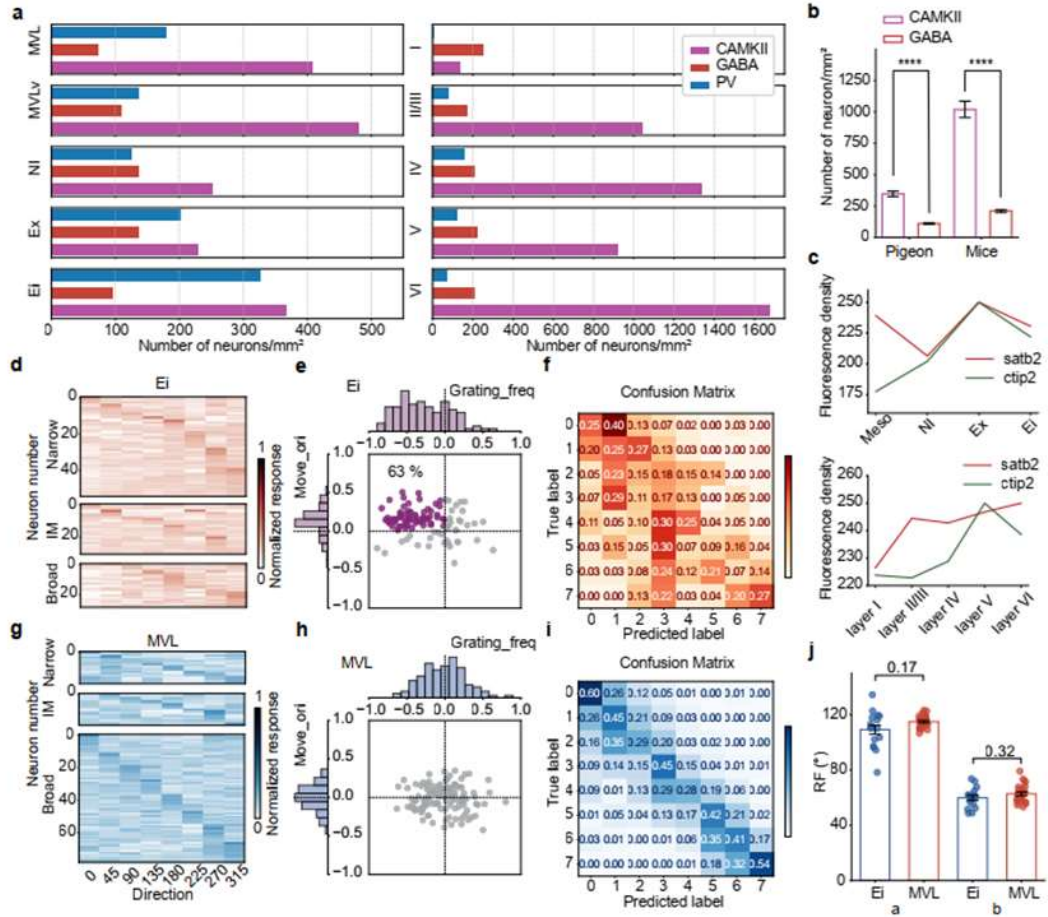

**Fig. 2. Hierarchical Organization of Neurons in the Visual DVR and Their Roles in Visual Information Processing. a**, Density of CaMKII+, GABA+, and PV+ neurons across different layers of the mice VISp ($n = 16$) and across distinccccct nuclei within the pigeon visual DVR ($n = 8$). **b**, Comparison of excitatory and inhibitory neuron densities between mouse VISp and the pigeon visual DVR. unpaired $t$-test: $t = 12.35, 8.704$, $p < 0.0001$, Bars represent mean ± SEM across populations. **c**, Fluorescence density profiles of Satb2 and Ctip2 immunoreactivity in visual DVR (top, $n = 3$) and rat VISp (bottom, $n = 4$). **d**, Directional tuning of Ei neurons in response to motion stimuli. **e**, Joint selectivity of Ei neurons for motion direction and spatial frequency of gratings. **f**, Confusion matrix of natural image classification performance by Ei neurons. **g**, Directional tuning of MVL neurons in response to motion stimuli. **h**, Joint selectivity of MVL neurons for motion direction and spatial frequency. **i**, Confusion matrix of natural image classification by MVL neurons. **j**, Comparison of receptive field sizes between Ei and MVL neurons (Ei: 109 ± 16°, 60 ± 9°; MVL: 115 ± 4°, 63 ± 6°; Wilcoxon rank-sum test).

Furthermore, Satb2 was mainly expressed in excitatory neurons in layers II-V, while Ctip2 (also known as Bcl11b) was predominantly expressed in layer V-VI cortical neurons, which are associated with cortical output functions in the mammalian cortex(Alcamo et al., 2008; Nikouei et al., 2016). Satb2 is also considered a marker gene for the pigeon mesopallium(Zaremba et al., 2025). Our quantitative fluorescence intensity results demonstrated that Satb2 expression was highest in the MVL region, with almost no Ctip2 expression, and both Satb2 and Ctip2 were relatively high in Ento. In comparison to VISp, Satb2 expression was strongest in layer II/III, while Ctip2 was concentrated in layer V (Figure 2c, Supplementary Figures 2a, b). These results suggest that the MVL region shares molecular marker expression and potential functional similarities with layers II/III of the mammalian cortex, possibly involved in information integration and transmission, while Ei exhibits molecular characteristics resembling those of layer V.

Next, electrophysiological recordings in response to simple visual stimuli were analyzed to examine the potential hierarchical organization within the visual DVR region (Methods). In terms of simple visual stimulus processing, we first found statistically significant differences, with Ei neuronal population showing a strong preference for specific stimulus categories (Figures 2d-e, Supplementary Figures 12d-e), whereas MVL neuronal population exhibited a more balanced response within the same stimulus category (Figures 2g-h, Supplementary Figures 12d-e). As for complex visual stimulus processing, the MVL's average classification accuracy was 43%, demonstrating strong natural image category selectivity (Figure 2i), while Ei showed no category selectivity (average classification accuracy 18.75%, random guess accuracy 12.5%, Figure 2f).

Moreover, neurons in functional columns typically process information from the same visual field, forming consistent receptive field (RF) structures in the mammalian visual cortex(Garrett et al., 2014; Murgas et al., 2020; Siegle et al., 2021). To verify this in MVL and Ei, we recorded receptive fields using 3.6° four-color block stimuli in anesthetized pigeons (3 pigeons, 16 neurons). By fitting the average receptive fields as ellipses (Supplementary Figure 12a), we found no significant statistical differences between the average receptive field diameters of the MVL and Ei regions (Figure 2j; long axis: p=0.17; short axis: p=0.32, Wilcoxon rank-sum test). These findings further support that the vertical spatial distribution of the MVL and Ei regions may reflect a columnar organizational structure similar to that of the mammalian cortex.

In conclusion, similar to the mammalian cortex, the number of excitatory neurons in the DVR visual region was greater than that of inhibitory neurons. PV+ neuron density in Ei was high, resembling layer 4. CaMKII + neuron density in the MVL region was also relatively high. Based on the expression patterns of Satb2 and Ctip2, MVL appears to correspond to cortical layers II/III, whereas Ei aligns with layer V. Furthermore, Ei and MVL shared similar receptive field structures, and there existed a hierarchical processing of visual information in the visual DVR region, from lower to higher levels.

**2. The MVL serves as the primary target for receiving projections from the E nucleus.**

To precisely delineate MVL, we employed HSV-H129-4G, comparing its efficacy with conventional retrograde tracer CTB and CO. Each of those was injected to MVL or Ei regions unilaterally (Methods). HSV tracing revealed that the labeled extent of MVL expanded progressively with longer infection durations (Fig. 3b, c, f). At all HSV infection durations, the region of MVL defined by HSV was all significantly larger than those defined by CO and CTB (n = 8) (Fig.3e, Supplementary Figure 5a). Cell densities within MVL at 46, 56, and 72 hpi showed no significant differences (Fig. 3e, Supplementary Fig. 3c–e, Supplementary Fig. 5a).

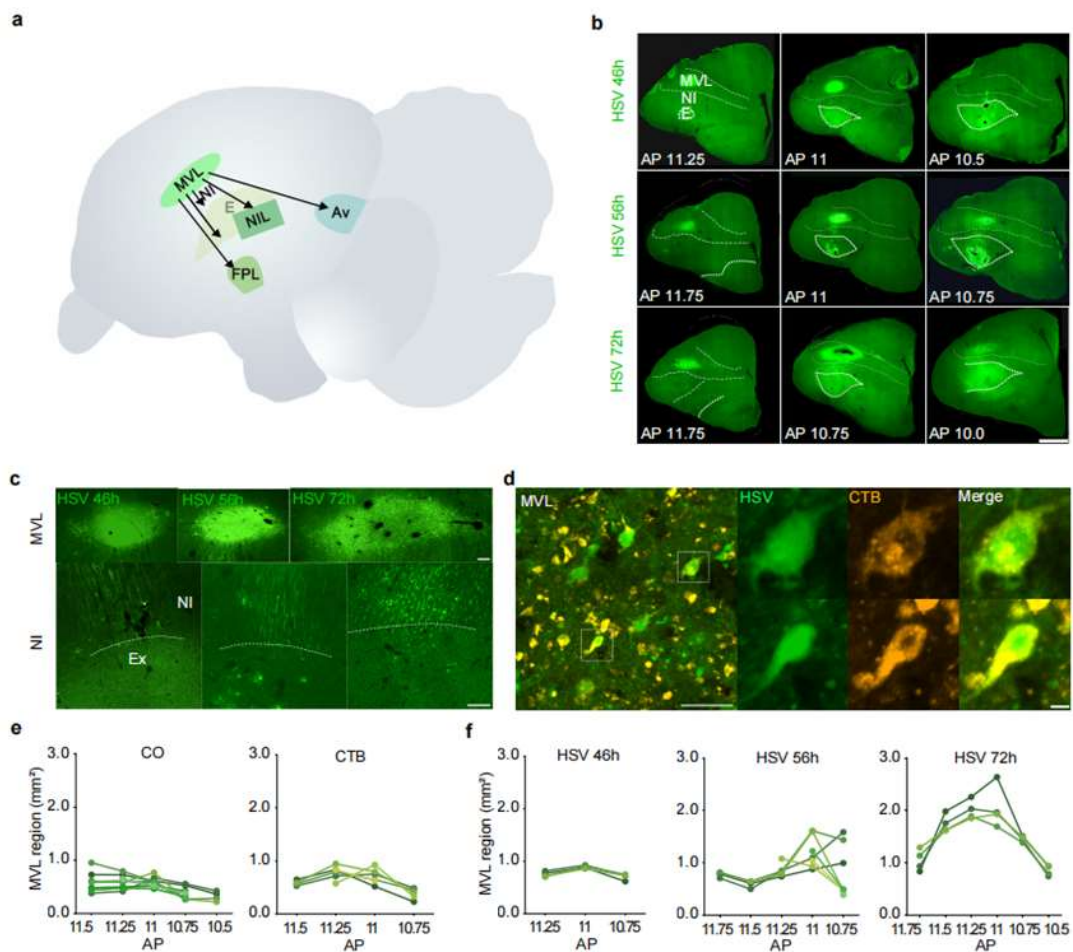

**Fig. 3. HSV-H129-4G Infection Patterns Following Injections into the E and MVL Regions. a**, Projection targets of MVL following a 56-hour HSV injection into MVL. **b**, Distribution of HSV infection in the MVL region following Ei injections at 46 h, 56 h, and 72 h post-infection. Injection sites and infected areas are shown. Scale bar = 2000 μm. **c**, Comparison of the spread of infection in the MVL and Ni regions at three post-

infection time points. Scale bar = 200 μm (both panels). **d**, Co-labeling results in MVL following dual injections of CTB and HSV into the same Ei site (n = 3). Scale bar = 50 μm. **e**, Area size of MVL labeling following cytochrome oxidase staining, CTB and HSV injections. Individual data points are overlaid for each group (n = 4). **f**, Quantified MVL labeling area (mm²) from HSV injections into the Ei region at three time points post-infection (n = 3). **Abbreviations:** NI = nidopallium intermediale, FPL = fasciculus prosencephali lateralis, Av = acropallium, pars ventralis, NIL = nidopallium intermedium lateralis, AP = anteroposterior.

We found that projections between Ento and MVL occurred earlier than other regions. Previous studies have suggested that both the MVL and NI are downstream brain regions of the Ento(Atoji and Wild, 2012; Krützfeldt and Wild, 2005). However, at 46 hours post-HSV infection, positively labeled neurons were observed exclusively in the MVL region across the whole brain. (Fig. 3b) At 56 hpi, the infection had expanded in MVL, and NI showed sparse labeling (unpaired t-test, p < 0.0001, n = 6) (Fig. 3c, Supplementary Fig. 3g). Additionally, at 56 hpi, fluorescent signals were also observed in the NCL (nidopallium caudolaterale), lateral Ad (dorsal arcopallium), part of Av (ventral arcopallium), and bilaterally Rt (nucleus rotundas thalamic) (Supplementary Fig. 5b). In contrast, CTB injection into Ei resulted in labeling in both MVL and NI, and labeling in the ipsilateral DMP (Nucleus dorsomedialis posterior thalami), AL (ansa lenticularis), and Rt, which is consistent with previous reports(Atoji and Wild, 2012; Krützfeldt and Wild, 2005), but no labeling was detected in contralateral Rt (Supplementary Figs. 5a, 6a).

Following HSV injection of in MVL, prominent labeling was observed in both Ei, Ex and NI at 56 hours post-infection (Supplementary Fig. 3a). At 72 hours, labeling extended to HA (hyperpallium apicale), with higher infection density in dorsal mesopallium compared to the ventral area where MVL was located. In the nidopallium, sparse labeling near the caudal Ento was more concentrated in Ei (Supplementary Fig. 3b). Based on our CTB injection results and previously reported MVL projection patterns(Atoji and Wild, 2012), we propose that within 56 to no more than 72 hours post-HSV infection, neuronal projections remained confined to secondary-order nuclei. (Fig. 3a, Supplementary Fig. 6b). Plus, co-labeling of neurons (HSV+CTB+) was observed in MVL (Fig. 3d,method), indicating that the MVL to Ei form a reciprocal connection.

Additionally, we utilized the VSV-GFP (another anterograde polysynaptic tracer) in pigeons (n = 5). Even after 3 days post-infection, this virus failed to traverse to secondary nuclei. However, following injections into the Ei or MVL regions, the fiber trajectories from Ei to MVL mirrored those observed with HSV, and the projection fibers from MVL can terminate in Ei, displaying parallel arrangements and distinct boundaries within MVL.

These results indicate that, MVL receives inputs from Ei primarily, and these projections are predominantly ipsilateral, whereas projections from Ei to Rt are bilaterally distributed.

### 3. Excitatory PV+ neurons play a critical role in the Ento-MVL circuit.

We found that, the majority of MVL neurons receiving projections from Ei were excitatory CaMKII-positive in HSV 56 hpi (one-way ANOVA, p = 0.0038) (Fig. 4a, c, d). Similarly, CTB tracing results revealed that most double-labeled neurons were CaMKII-positive (one-way ANOVA, p < 0.0001), while rarely a small proportion were GABAergic in MVL (one-way ANOVA, p < 0.0001) (Supplementary Fig. 9c, f). Similarly, spike waveform analysis of single MVL neurons revealed a significantly higher proportion of broad-spiking (BS) neurons compared to intermediate-spiking (IMS) and narrow-spiking (NS) neurons. (Fig. 4e). These findings indicate that both input and output neurons in MVL are predominantly excitatory. Moreover, we also observed that HSV-labeled fibers from Ento to MVL expressed PV, suggesting that Ento projection neurons targeting MVL express PV (Fig. 4b, f).

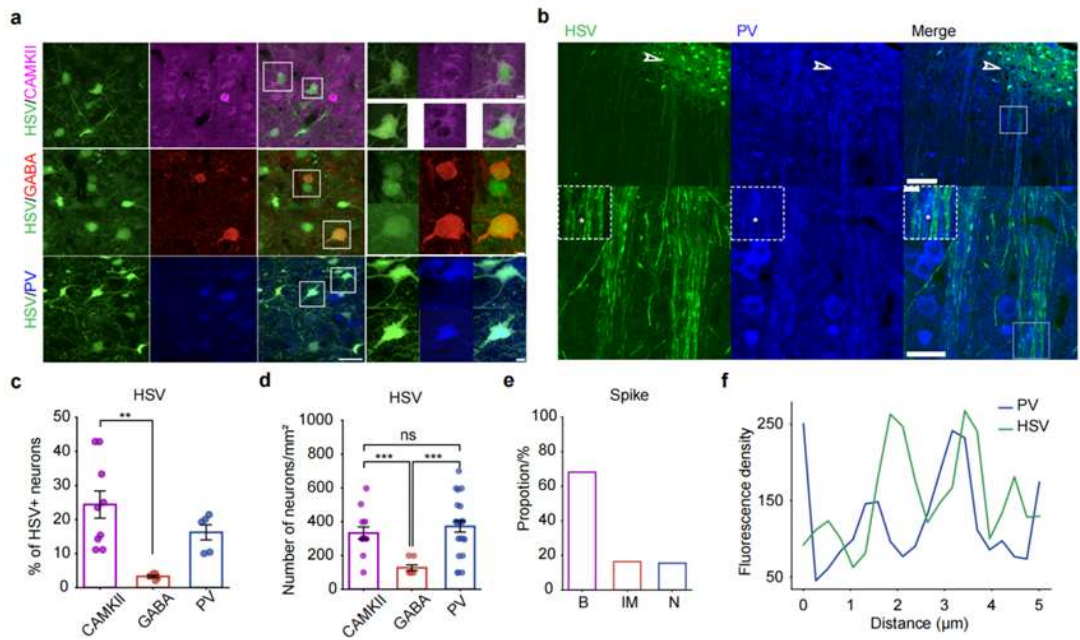

**Fig. 4. Preliminary Classification of Projection Neurons Between Ento and MVL. a**, Representative images of HSV+ in MVL at 56 hours after HSV injection in Ei. Scale bars = 30 μm, 5 μm in right panel. **b**, Co-localization of HSV+ and PV+ fibers (projecting from Ei) in MVL. Scale bars = 100 μm in upper panel, 20 μm in low panel, 5 μm in inserts. **c**, Proportion of different neuron types in MVL following HSV injection in Ei (56 h post-infection): CamKII group (n = 9), GABA group (n = 5), PV group (n = 5). Ordinary one-way ANOVA, F = 8.041, p = 0.0038. **d**, Density of different neuron types in MVL at 56 h post HSV injection in Ei. Ordinary one-way ANOVA, F = 23.66, p < 0.001. **e**, Proportions of three spike types recorded by electrophysiology. **f**, Fluorescence density profiles of PV- and HSV-labeled samples plotted against distance (μm), illustrating comparative distribution patterns in MVL.

In mice, excitatory PV+ neurons have been identified in layers V and VI of VISp (Supplementary Fig. 7a–c). In MVL, the density of excitatory PV+ neurons was significantly higher than inhibitory PV+ neurons (unpaired t-test, p = 0.0034) (Fig. 5a, c, d). Likewise, in Ei, the density of excitatory PV+ neurons was significantly higher than that of inhibitory PV+ neurons (unpaired t-test, p < 0.0001) (Fig. 5b, d). Notably, among MVL PV+ neurons receiving input from Ei, the majority were also excitatory (unpaired t-test, p < 0.0001) (Fig. 5b, d).

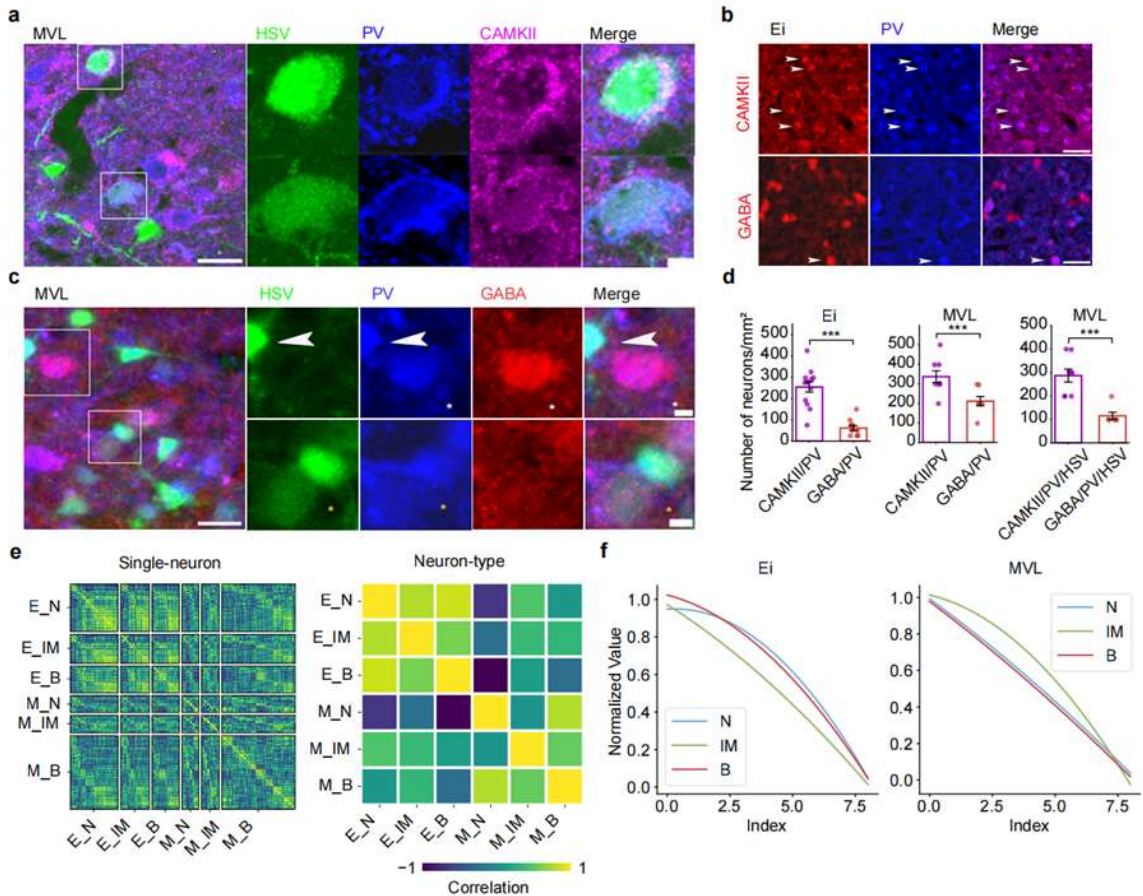

**Fig. 5. Proportions of Excitatory and Inhibitory Components in PV+Neurons in MVL and Ento, and Their Functional Differences in Visual Processing. a,** Representative images showing HSV+PV+CamKII+ neurons in MVL. Scale bars = 20 μm, 5 μm in right panel. b, Representative images of PV+CamKII+ and PV+GABA+neurons in Ei. Scale bar = 50 μm. c, Representative images showing HSV+PV+GABA+neurons in MVL. Scale bars = 20 μm, 5 μm, 5 μm in right panel. d, Comparison of densities between PV+CamKII+ and PV+GABA+neurons in Ei and MVL, and between HSV+PV$^+$/CamKII+and HSV+PV+GABA+neurons. $N$ = 12, unpaired $t$-test: $t$ = 8.349; $N$ = 9, unpaired $t$-test: $t$ = 3.471, $p$ = 0.0034; $N$ = 9, unpaired $t$-test: $t$ = 5.785, $p$ < 0.0001. e, Similarity map of response profiles between single neurons and neuron types. Naming convention: nucleus_neuron type. f, Fitted tuning curves of neuronal selectivity for different neuron types in Ei and MVL.

To further investigate the properties of neurons in the Ento–MVL circuit, we compared the tuning similarity of neurons with different spike widths based on electrophysiological recordings. At the single-neuron level, neurons in both Ei and MVL exhibited distinct tuning profiles (Fig. 2d, g; Supplementary Fig. 13a, b; 14a, b). Similarity analysis revealed that neurons of both the same and different types, within or across regions, could show either highly similar or highly dissimilar tuning (Fig. 5e). At the population level, similarity analyses showed that within the same region, NS neurons shared greater similarity in their population responses with BS neurons (Fig. 5e; Supplementary Fig. 13c, 14c). Curve fitting of population selectivity further confirmed that NS and BS neurons exhibited similar population-level tuning (Fig. 5f; Supplementary Fig. 13, 14). These findings suggest that putative PV+ neurons (NS) and putative excitatory neurons (BS) within the same region share similar tuning properties, providing additional support for the hypothesis that PV+ neurons may exhibit excitatory functions in Ento-MVL circuit.

Overall, PV+ neurons in the avian brain may have excitatory functions and exhibit tuning properties similar to those of classical excitatory neurons.

## 4. Distinct Processing Patterns of the Ento–MVL Circuit for Moving versus Looming Stimuli

After characterizing both the projections and intrinsic properties of the Ento-MVL circuit, we next examined its functional response profiles. Previous studies have demonstrated that avian visual pathways process moving objects with remarkable speed (Hataji et al., 2019; Hu et al., 2025). Accordingly, we classified our visual stimuli into two categories—looming and moving—and compared how different neuronal subtypes within the Ento-MVL circuit respond to these presentations (Fig. 6a). Our results revealed that, in terms of firing intensity, the NS and IMS neurons in the E nucleus exhibited very strong responses to moving stimuli, while the BS neurons also fired robustly under the same condition. By contrast, within the MVL region, looming stimuli evoked stronger firing across all three neuronal types compared to moving stimuli. Regarding response duration, although looming and moving stimuli were presented for identical durations, all neuronal types maintained prolonged firing under moving stimulation. In terms of response delay, moving stimuli induced a faster achievement of peak firing rates across every neuronal class (Fig. 6b). Together, these findings indicated that the Ento-MVL circuit responded to moving objects with both extended firing durations and accelerated peak latencies, manifesting as a rapid yet sustained activation profile. Among all subtypes, the putative PV neurons (NS) exhibited the fastest response delay and the strongest firing rate.

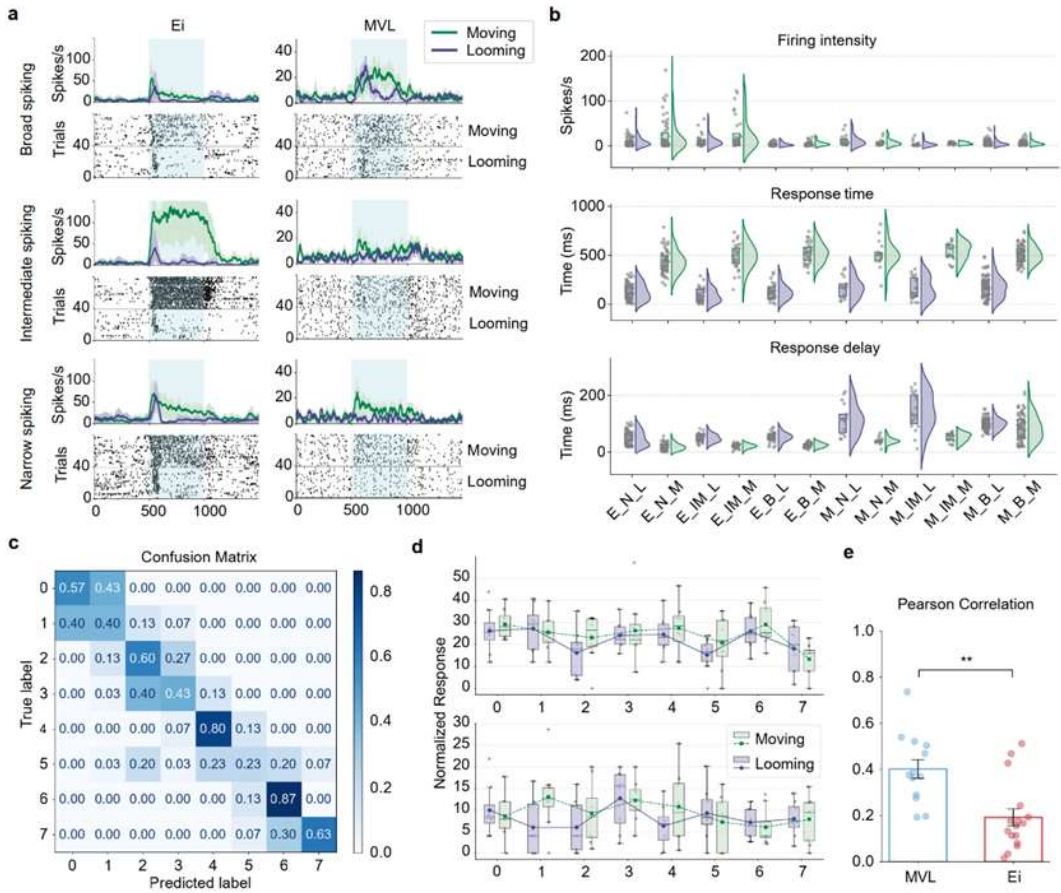

**Fig. 6. Distinct Firing and Information Processing Patterns in MVL and Ei in Response to Looming and Moving Stimuli. a**, Representative firing rate curves and raster plots of neuronal responses under looming and moving stimulus conditions. **b**, Boxplots and violin plots showing the distributions of firing intensity, response duration, and response delay for neurons under both stimulus types. **c**, Confusion matrix illustrating MVL classification performance for moving natural images. **d**, Neuronal selectivity to natural images under looming versus moving presentations. The upper panel shows examples with similar selectivity across conditions, while the lower panel shows examples with large selectivity differences. **e**, Statistical comparison of selectivity similarity between looming and moving conditions in MVL and Ei neurons. Wilcoxon Rank-Sum Test, $p < 0.01$.

To further investigate pigeons' processing of moving objects, we recorded neuronal responses while pigeons viewed natural images presented in either looming or moving formats. In assessing MVL image-classification performance, we found that moving presentations of natural images yielded higher classification accuracy in MVL (Fig. 6c), with an average accuracy of 56.63%. Next, we examined how individual neurons category selectivity (as reflected by their firing intensity distributions) differed between looming and moving presentations. Some neurons exhibited highly similar selectivity profiles across the two modes, whereas others showed marked differences (Fig. 6d). By computing Pearson correlation coefficients between each neuron's category-response vectors under looming versus moving conditions, we determined that neurons in the MVL region displayed significantly higher correlation (Fig. 6e).

In summary, during moving presentations, the extended firing durations and accelerated peak latencies of neurons (especially NS) within the Ento-MVL circuit, together with MVL's superior classification performance and increased correlation of neuronal response patterns, indicate that the Ento-MVL circuit is uniquely optimized for processing moving-object information—an adaptation that may be critical for navigating complex three-dimensional visual environments.

## 5. HS-RNN Model Simulating Information Processing in the E-MVL Circuit

Building on our prior results, we implemented a recurrent neural network (RNN), which is a widely adopted framework for modeling biological neural connectivity and computational mechanisms (Hasani et al., 2022; Koutník et al., 2014; Xiao et al., 2024). Using this framework, we created a heterogeneous-speed RNN (HS-RNN) model that captured neurons with diverse firing rates and thus emulates the Ento–MVL connectivity and computation (Methods). The network consisted of two layers, E (layer 1) and MVL (layer 2) (Fig. 7a). To mimic the narrow and broad spiking behaviors of distinct neuron types, we divided the hidden-layer neurons into three groups (S1, S2, and S3) based on electrophysiological clustering; each group propagated signals at a different speed, producing asynchrony in downstream transmission (Fig. 7b). We trained this model on randomly generated looming and moving sequences paired with the corresponding neuronal population responses. To further simulate the visual perceptual decision process, we also evaluated its performance on an image classification task by coupling an image encoder with the HS-RNN (Methods).

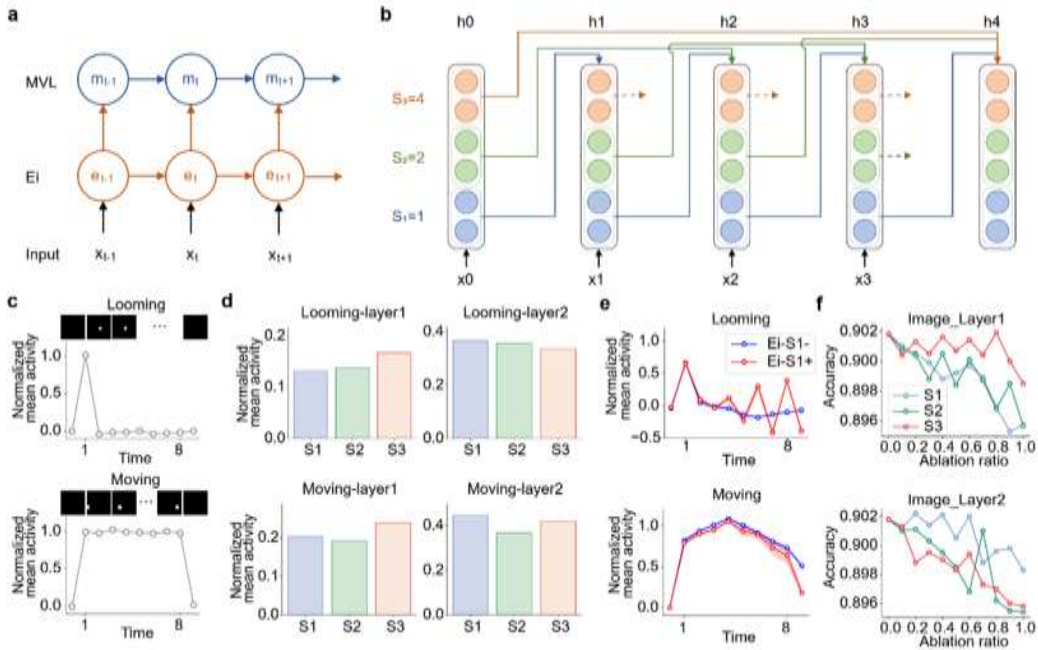

**Fig. 7. Architecture of the Heterogeneous-Speed RNN (HS-RNN) and Experimental Results. a**, Overall structure of the HS-RNN model, consisting of two layers: E (layer 1) and MVL (layer 2). **b**, Parameter update strategy within the same layer. Neurons are divided into three groups—S1 (NS), S2 (IMS), and S3 (BS)—each influencing the network at different update speeds. **c**, Normalized network output activity in response to looming and moving stimuli. **d**, Average activation intensity of different types of hidden neurons under looming and moving stimulus paradigms. **e**, Effects of ablating excitatory versus inhibitory S1 neurons in the E layer on network output. **f**, Effects of ablating different neuron types with varying ablation ratios on natural image classification accuracy.

After training, the model's outputs closely recapitulated the neural response profiles evoked by moving and looming stimuli (Fig. 7c). To quantify the contributions of each spiking-speed subgroup, we computed the mean activity of hidden-layer neurons in both E and MVL layers under each stimulus condition. Under moving stimulation, all three groups exhibited elevated mean firing rates, with the fast-spiking neurons (S1) showing the most pronounced increase (Fig. 7d). This finding indicates that fast-spiking neurons preferentially participate in motion processing. To investigate the roles of different cell types, we performed targeted ablations of specific neuron subtypes in designated layers. Ablating S1 (fastest spiking) neurons in the E layer caused the most severe degradation in overall network output, while in the MVL layer, the greatest impairment was observed when ablating S3 (slowest spiking) neurons (Supplementary Fig. 15a, b). Furthermore, we selectively ablated excitatory (positive-weight) and inhibitory (negative-weight) S1 neurons in both the E and MVL layers. The removal of excitatory fast-spiking neurons resulted in greater performance deficits than the removal of inhibitory ones (Fig. 7e;

Supplementary Fig. 15c), highlighting the critical role of excitatory fast-spiking neurons in basic information processing within the Ento-MVL circuit.

To evaluate the network's capacity for higher-level visual tasks, we trained the HS-RNN on the CIFAR-10 image-classification benchmark. After training, we applied a graded series of ablation coefficients to each neuron type in both layers and measured the resulting changes in classification accuracy. In E layer, ablating S1 and S2 neurons caused the most severe performance degradation. In MVL layer, ablating S3 and S2 neurons induced the greatest accuracy loss (Fig. 7f). These patterns suggest that NS plus IMS neurons in E and BS plus IMS neurons in MVL are the principal contributors to image processing.

Overall, the HS-RNN findings validate our electrophysiological and immunohistochemical observations. They further imply that positive PV+ neurons, which are characterized by narrow and fast spiking, endow the pigeon Ento-MVL circuit with rapid motion processing capabilities, while slow-spiking (largely excitatory) neurons contribute more substantially to complex information processing and object recognition.

## Discussion

In our study, we discovered a reciprocal and preferential neural circuit between the Ento and MVL in pigeons. This circuit is predominantly mediated by excitatory PV+ neurons and shows significant parallels to the excitatory PV+ neurons in SC. We propose that the visual-related DVR do not strictly adhere to the "nuclear-layer" model:MVL and Ei contain neurons resembling those of deep neocortical layers, alongside populations implicated in superficial layer–like sensory processing.

At the cellular level, our results show that both MVL and Ento are rich in excitatory PV+ neurons (Figure 5a, b, d). Functionally, PV+ neurons in MVL and Ei display strong discharge activity in response to looming visual stimuli (Figure 6a, Supplementary Figures 7c, d, e, f), resembling the physiological function of PV+ neurons in the superficial gray layer (suG) of the SC, which mediate fear-avoidance responses(Shang et al., 2015). Additionally, the connections between MVL and the FPL reveal its critical role in visual-motion integration (Figure 3a). Based on the gene expression patterns of layer-specific markers, the co-expression of SATB2 and CTIP2 in Ento indicate that Ento is corresponded with mammalian layer V(Figure 2c). Recent single-cell transcriptomics further show that certain neuronal types in Ento and mesopallium share significant transcriptional similarities with the claustrum and piriform cortex of mammals(Zaremba et al., 2025), supporting our hypothesis that this circuit exhibits features analogous to mammalian deep-layer or subcortical functional circuits.

Moreover, the Ento–MVL circuit also manifests a hierarchical processing structure for visual information, functionally analogous to the processing mode between layer IV and layer II/III in the mammalian neocortex: MVL integrates diverse low-level visual features originating from Ei, thereby minimizing reliance on individual features while maintaining high selectivity in the processing of natural image categories.(Figure 2f). Immunofluorescence labeling shows that the expression of hierarchical marker genes in MVL aligns highly with layer II/III of the rat neocortex, indicating functional hierarchical differentiation similar to that of the neocortex (Figure 2c). Furthermore, HSV tracing results demonstrate that MVL has dense inter-cortical connections, and all its output projection neurons are excitatory, with superior recognition of motion images dependent on the activity of BS neurons (Figures 7d, Supplementary Figures 3b). This further supports its functional analogy to layers II/III of the mammalian neocortex. In addition, our HS-RNN results confirmed that hierarchical information processing in the Ento-MVL pathway is mediated by excitatory PV+ neurons in Ei.

Based on these two sets of results, we propose that the Ento–MVL circuit not only resembles the structural-functional organization of mammalian midbrain integration pathways but also exhibits a high degree of consistency with the superficial perceptual processing modules of the neocortex. Consistent with our findings, another nucleus in DVR, HVC, also shows transcriptomic features corresponding to both deep and superficial cortical layers in mammals, reinforcing our perspective that a single nucleus within the avian DVR can concurrently perform multiple functions analogous to distinct cortical layers in mammals(Colquitt et al., 2021).

In contrast to mammals, the neural connections between MVL and Ento are restricted to the same hemisphere. Ei receives projections from ipsilateral Rt neurons and provides feedback to both ipsilateral and contralateral Rt regions (Supplementary Figures 4c). In mammals, layer II/III neurons in V1 regulate interhemispheric visual communication via inhibitory PV+ neurons in the corpus callosum(Wang et al., 2023). We hypothesize that this difference may be related to the avian trait of unihemispheric sleep. Studies have shown that raptors maintain vigilance during long flights by employing unihemispheric sleep to meet environmental demands(Miñano et al., 2023). Similarly, monocular deprivation in pigeons leads to differences in the activity frequency between the two hemispheres of the midbrain, suggesting that avian unihemispheric sleep relies on the independent processing of visual information by each hemisphere(Rattenborg et al., 2019). Projection from the Rt to Ento facilitates the convergence of binocular inputs, indicating that the left Ei has the capacity to process binocular visual information.Therefore, we speculate that the balance of binocular visual input in pigeons may be achieved at the midbrain level, obviating the need for extensive interhemispheric communication for binocular vision(Supplementary Figures 4c).

In comparison to previous studies on avian visual neural circuits, our study further clarifies the projection priority between the Ei and MVL. For the first time, we used viral tracing using HSV, VSV (vesicular stomatitis virus), AAV1 (adeno-associated virus type 1), and adenovirus (AD) to study the neural connectivity. After two months of AAV1 infection, labeled cells were only observed in the Ei, with no noticeable projection fibers detected. AD injections resulted in nonspecific labeling, presenting as diffuse high-density fluorescence throughout the brain, likely due to the strong immune response of pigeon. VSV virus infection clearly labeled projection fibers near the injection site, but failed to infect secondary neurons, with fluorescence confined to fibers in secondary regions (Supplementary Figures 4a, b).

Although this study provides preliminary insights into the functional characteristics of excitatory PV+ neurons in the Ento-MVL pathway of avian visual circuits, it has certain limitations. Due to the strong immune response of pigeon, some viral vectors (such as AAV1 and AD) failed to achieve effective tracing, potentially resulting in insufficient identification of sparse projection pathways. In future studies, optimizing viral tools could enhance the specificity and sensitivity of tracing. Furthermore, while our study speculated on the functional types of PV+ neurons based on electrophysiological recordings and HS-RNN analysis, verification at the single-cell isolation and full transcriptome level is still lacking. Future research should integrate transcriptomic sequencing to further elucidate the classification and function of excitatory PV+ neurons. Functionally, this study primarily focused on neural responses to approaching and motion visual stimuli, with limited systematic investigation into static scene processing, spatial attention regulation, and multimodal integration. Future studies could use virtual reality (VR) technology to more deeply analyze the encoding characteristics of the Ento-MVL pathway in natural visual behaviors under free behavioral conditions.

## Materials and Methods

## 1. Animal and surgeries

### 1.1 Pigeon Handling

Adult male and female pigeons (Columba livia) aged 1.5 to 2 years and weighing 350–450 g were used in this study.   In addition, male C57BL/6J mice (7 weeks old) and male Sprague-Dawley rats (8 weeks old) were also used.Prior to the experiments, the pigeons were housed in a temperature- and humidity-controlled environment for at least two weeks to acclimate, under a natural light-dark cycle. Water and food were provided ad libitum. Following surgery, the pigeons were individually housed for the duration of the experiments.

All experimental procedures involving live animals and brain tissues adhered to the guidelines of Zhengzhou University and Beijing Normal University. The protocols were approved by the Institutional Review Board (IRB) of Zhengzhou University and the Institutional Animal Care and Use Committee (IACUC) of Beijing Normal University.

### 1.2. Neural Tracer Injection

Before the experiment, the pigeons were anesthetized with 20% ethyl carbamate, with the dosage calculated based on body weight (0.3 mL/100 g). After complete anesthesia was achieved, the experiment began. The pigeons were fixed in a stereotaxic instrument (RWD Life Science) to ensure head stability. The anterior-posterior (AP) zero point was determined by measuring the distance between the Bregma point and the ear bar. The specific locations of Ei and MVL were referenced from the 1976 pigeon brain atlas (MVL: AP 11, ML 4.5, DV 2.0; Ei: AP 10.75, ML 5.3, DV 4.7)(Karten and Hodos, 1967). A microinjection syringe (RWD Life Science) was used to slowly inject the required virus or tracer substances (HSV-H129-4G, GeneChem; AAV1-CAG-chR2-GFP, OBIO; VSV-EYFP, Brain Case; CTB-555, Brain Case). The vertical depth (DV) zero point was defined when the tip of the injection needle made contact with the dura mater. After the injection, the needle was held in place for 5 to 10 minutes before being slowly removed, then the wound was sutured and treated with erythromycin ointment. The pigeons were returned to their cages after recovery. The infection times for HSV virus were 46 hours, 56 hours, 72 hours, and 3 days; AAV1 infection time was 2 months; VSV's infection time was 72 hours; CTB infection time was 7 days. It is important to note that, in the co-labeling experiment with HSV and CTB, after CTB was injected, HSV was injected into the same side of Ei 7 days later.

## 1.3 Brain Tissue Processing

After the virus or tracer infection period, perfusion and brain extraction were performed. Pigeons were anesthetized with an overdose of 20% ethyl carbamate (Aladdin, U108629) at a dose of 0.5 mL/100 g. Once fully anesthetized, the chest cavity was opened with scissors to expose the heart. A perfusion needle was inserted into the left ventricle, and the right atrium was quickly and accurately incised. The pigeons were first perfused with physiological saline at a steady rate until the effluent from the right atrium became clear, followed by perfusion with 4% paraformaldehyde. After perfusion, the brain tissue was carefully extracted using bone forceps and fixed in 4% paraformaldehyde. Following fixation, the brain was dehydrated in 30% sucrose solution. The tissue was then sectioned into 50-μm-thick slices using a cryostat and collected into a 24-well plate containing 0.1 M PBS buffer for subsequent experiments.

## 1.4 Immunofluorescence

Frozen brain sections were mounted onto slides and air-dried. A PAP pen (Abcam, ab2601) was used to draw a circle around the staining area. The sections were washed three times with 0.01 M PBS (pH 7.4) for 5 minutes each. They were then blocked at room temperature for 40 minutes with 5% donkey serum or goat serum (YTHX Biotechnology, Beijing, China) diluted in 0.5% Triton X-100/PBS. The following primary antibodies were applied overnight at 4°C: Rabbit anti-CaMKII (Abcam, ab216503, 1:200), Mouse anti-PV (Chemicon, MAB1572, 1:500), Rabbit anti-GABA (Abcam, A2052, 1:200), Mouse anti-Satb1 + Satb2 (Abcam, ab51502, 1:100), Rat anti-Ctip2 monoclonal antibody (Abcam, ab18465). Primary antibodies were diluted in the same serum solution used for blocking. The next day, the slides were equilibrated at room temperature for 2 hours in a humidified chamber, washed four times with 0.01 M PBS, and incubated with the following secondary antibodies at room temperature for 2 hours: Alexa Fluor 647 donkey anti-rabbit (1:200, Jackson, cat 711-606-15), Alexa Fluor 594 donkey anti-mouse (1:500, Abcam, AB150013), Alexa Fluor 405 donkey anti-mouse (1:100, Abcam, ab175658), Alexa Fluor 488 goat anti-mouse (1:500, Abcam, AB150117), Alexa Fluor 488 donkey anti-rat (1:500, Abcam, ab150153). The secondary antibodies were diluted in 0.01 M PBS. Following incubation, the sections were washed six times with 0.01 M PBS and mounted with anti-fade mounting medium under coverslips.

## 1.5 Cytochrome Oxidase Staining

Frozen brain sections were washed three times with 0.01 M PBS (pH 7.4) for 5 minutes each. Cytochrome oxidase (CO) clearing solution (labeled with benzidine, Solarbio, G2410) was applied to cover the entire slide for 2 minutes, then removed. The sections were incubated in DAB solution at 37°C for 1 hour, followed by rinsing with distilled water. Dehydration was performed using absolute ethanol for 5 minutes, and the slides were cleared with xylene for 5 minutes. Finally, the slides were mounted with neutral resin.

## 1.6 Electrode Implantation Surgery

The electrode implantation procedure was similar to the stereotaxic injection process. After exposing the skull and determining the DV zero point, the dura mater at the implantation site was removed. Electrodes were slowly implanted at a rate of 10 minutes per millimeter. Once the electrodes were in place, they were fixed using dental cement. The pigeons were allowed to recover for 7 days before signal acquisition.

## 1.7 Image Acquisition and Processing

Prepared slides were imaged using a ZEISS inverted fluorescence microscope (AxioCam MRm) and a ZEISS laser confocal microscope (LSM700). Whole-brain sections were scanned using a 20× objective, and specific brain regions were imaged using a 63× objective. The acquired images were processed with ZEN 3.0 (Blue Edition) and ZEN 2009 Light Edition software for layer stacking and channel merging. Cell counting and fluorescence intensity analysis were performed using Fiji software. Statistical analyses were conducted using GraphPad Prism 9.

## 2. Electrophysiological Signal Acquisition System

## 2.1 Apollo System

The Apollo electrophysiological signal acquisition system is a high-performance, multi-channel platform designed for advanced neuroscience and biomedical research. The core modules of the Apollo system include signal amplifiers, analog-to-digital converters (ADC), and efficient data processing units, which work collaboratively to ensure data accuracy and real-time processing. For this study, a 16-channel configuration was utilized with a sampling rate of 30 kHz.

## 2.2 Electrodes

Spike signals were recorded using KD-MEA electrodes manufactured by KedouBC. These 16-channel 4×4 array electrodes are composed of electrode wires with tapered tips made of tungsten, coated with a biocompatible insulating layer, and designed with sufficient rigidity. Different electrode configurations were customized based on the location and size of the brain regions: electrodes with a wire length of 6 mm and a spacing of 500 μm were used for the Ei region, while electrodes with a wire length of 4 mm and a spacing of 300 μm were employed for the MVL region.

## 2.3 Visual System

Visual stimuli were presented to the pigeon's right eye (Verhaal et al., 2012) using an LCD monitor with a resolution of 3840×2160 pixels and a refresh rate of 100 Hz. The left eye was occluded with a light-blocking strip. The display plane was aligned with the tangent point of a sphere centered on the right eye, with the monitor vertically fixed at a 29° angle relative to the pigeon's body midline. The distance from the center of the monitor to the right eye was 20 cm. The monitor horizontally covered 138° of the visual field and vertically covered 112°.

## 3. Visual Stimulation Paradigm

Visual stimuli were generated and controlled using the Psychophysics Toolbox in MATLAB (Brainard, 1997). For receptive field mapping, the stimulus consisted of small patches with a size of 3.6°, randomly displayed across the screen, covering the entire visual field. Each stimulus was presented for 500 ms with an interstimulus interval of 10 s.

For specific visual feature stimuli:

- **Color Stimuli**: Six colors—red, orange, yellow, green, cyan, and blue—were designed with incrementally increasing H values in the HSV color space, corresponding to progressively shorter wavelengths in the visible spectrum.
- **Motion Direction Stimuli**: Stimuli were presented at directions ranging from 0° to 315°, with 45° intervals.
- **Grating Frequency Stimuli**: Gratings with spatial frequencies of 0.01, 0.05, 0.1, 0.2, 0.5, 1, 2, 3, 4, and 5 cycles per degree were presented, representing increasing frequencies.
- **Natural Image**: We used a dataset comprising 8 distinct categories of natural images: *dog*, *elephant*, *flower*, *human*, *pigeon*, *ship*, *table*, and *tree*. For each category, 8 unique images were selected, totaling 64 images (Supplementary material).

All of the stimuli were presented fullscreen for 500 ms, with an interstimulus interval of 15 s. Each session was designed to present only one type of visual stimulus. Within a session, stimuli were presented in random order, and each stimulus was repeated 10 times.

## 4. Electrophysiological Signal Processing and Neuron Extraction

Raw data from different visual stimulation sessions of the same pigeon were concatenated, and their timestamps were aligned. This approach avoided the issue of neurons being difficult to align after sorting data from separate sessions independently. High-frequency signals were extracted using a band-pass filter (300–5000 Hz). Spike sorting was performed using Kilosort, and the output was manually curated using Phy, a graphical user interface for visualization(Pachitariu et al., 2016). Only well-isolated units with clear refractory periods in the autocorrelograms were selected for further analysis. After sorting, the spike times of each neuron were aligned with the visual stimulus timestamps. Neural signals from each trial were extracted, starting from the baseline period before stimulus onset to a short period after stimulus offset. For each neuron, 84 data points around the spike times were extracted from the filtered high-frequency signals and saved as the spike waveform for that neuron at that specific spike event. The average of all spike waveforms was calculated to represent the characteristic spike waveform of the neuron.

## 5. Neural Tuning Analysis

We used spike-triggered average (STA) analysis to examine the tuning of neuronal populations to different types of stimuli, which were grouped based on color (corresponding to visible light wavelengths), grating frequencies, and motion directions. Specifically, we first calculated the average neuronal response within a 50–300 ms time window after stimulus onset across 10 trials. These responses were then normalized into probability distributions. Neurons were sorted by their preferred stimuli, and response heatmaps were generated to visualize the tuning of individual neurons to the stimuli.

To measure the preference of different brain regions for specific stimulus types, we defined three selectivity indices: color_SI, motiondir_SI, and gratingfreq_SI.

$$Color_{SI} = \frac{S - L}{S + L} \tag{1}$$

Here, S represents short-wavelength stimuli, and L represents long-wavelength stimuli.

$$Motiondir_{SI} = \frac{FD - BU}{FD + BU} \tag{2}$$

Here, FD represents forward and downward motion, while BU represents backward and upward motion.

$$Gratingfreq_{SI} = \frac{H - L}{H + L} \qquad (3)$$

Here, H represents high-frequency gratings, and L represents low-frequency gratings.

These indices quantify the relative preference of neuronal populations for specific stimulus properties, with values closer to 1 or -1 indicating stronger selectivity, and values closer to 0 indicating more balanced responses.

## 6. Heterogeneous-Speed Recurrent Neural Network (HS-RNN)

### 6.1 Network Architecture

We propose a biologically inspired recurrent neural network architecture, termed Hetero-Speed RNN (HS-RNN), motivated by the e–MVL circuit in birds. HS-RNN captures time-dependent interactions among neuronal subtypes with different spike propagation speeds. The model consists of two stacked recurrent layers, with layer 1 representing the *E* nucleus and layer 2 representing the *MVL*. Each layer includes three types of recurrent units—fast ($s_1$), medium ($s_2$), and slow ($s_3$)—which update their hidden states at different temporal rates. This design mimics temporal integration observed in biological circuits and enables effective modeling of input sequences across multiple timescales.

### 6.2 Network Dynamics

At each time step $t$, the hidden state $h_t$ is updated as a weighted sum of past hidden states. Different neuronal subgroups access different temporal depths, reflecting their distinct integration timescales.

$$h_t = f\left(W_x x_t + W_h^{s_1} h_{t-t_1}^{s_1} + W_h^{s_2} h_{t-t_2}^{s_2} + W_h^{s_3} h_{t-t_3}^{s_3} + b\right) \qquad (4)$$

where $f(.)$ is a non-linear activation function (e.g., tanh), $h_{t-t_i}^{s_i}$ is the hidden state of the $s_i$ subpopulation at time $t - t_i$, $W_h^{s_i} \in \mathbb{R}^{N \times N_{s_i}}$ are trainable recurrent weights for each group, $W_x \in \mathbb{R}^{N \times d}$ is the input weight matrix, $b \in \mathbb{R}^N$ is the bias term.

Let $h_t^{(1)} \in \mathbb{R}^{N^{(1)}}$ be the hidden state of the first layer, and $h_t^{(2)} \in \mathbb{R}^{N^{(2)}}$ the second. The second layer receives input from the hidden state of the first layer:

$$h_t^{(2)} = f\left(W^{(2)} h_t^{(1)} + \sum_{i \in \{1,2,3\}} W^{(2,s_i)} h_{t-t_i}^{(2,s_i)} + b^{(2)}\right) \qquad (5)$$

### 6.3 Network Training

The model is trained on a sequence prediction task using the mean squared error (MSE) loss. The input to the network is a sequence of 10 grayscale images of size 8×8, where a white pixel indicates the presence of a target. Targets appear in two different stimulus types: *looming* and *moving*. The goal is to predict the corresponding neuronal responses. The predicted response at time step $t$ is denoted as $\hat{\mathbf{y}}_t$, given by:

$$\hat{y}_t = W_o h_t^{(2)} + b_o \tag{6}$$

The objective is to minimize:

$$\mathcal{L}_{MSE} = \frac{1}{T} \sum_{t=1}^{T} (\hat{y}_t - y_t)^2 \tag{7}$$

For the natural image sequence classification task, the model is trained using the cross-entropy loss. We use the CIFAR-10 dataset and simplify the task by employing a pretrained ResNet-18 as an image encoder to extract embedding vectors(He et al., 2016). These embeddings are fed into the HS-RNN, which predicts the image category. The predicted class probability at time step $t$ is denoted as $\hat{\mathbf{y}}_t$, given by:

$$\hat{y}_t = Softmax\left(W_o h_t^{(2)} + b_o\right) \tag{8}$$

The objective is to minimize:

$$\mathcal{L}_{CE} = -\frac{1}{n} \sum_{i=1}^{n} y_i{}^T \log(\hat{y}_i) \tag{9}$$

## 6.4 Network Ablation

To evaluate the contribution of each subcomponent within the HS-RNN architecture, we conducted a series of ablation studies by selectively modifying or removing key elements of the model. Each ablated variant was evaluated under the same conditions as the full model, using identical input sequences.

## Data availability

The data involved in this paper will be publicly available upon the publication of the paper.

## Code availability

Python code for HS-RNN model simulation from this work will be freely available online upon the publication of this paper.